# Modification of the zirconia ceramics by different calcium phosphate coatings: comparative study

A.I. Kozelskaya[a], E.N. Bolbasov[a], A.S. Golovkin[b], A.I. Mishanin[b],
A.N. Viknianshchuk[b], E.V. Shesterikov[a], A. Ashrafov[a], V.A. Novikov[c], S.I. Tverdokhlebov[a,*]

[a] *Tomsk Polytechnic University, 30 Lenin Avenue, Tomsk, Russian Federation*
[b] *Almazov Federal Medical Research Centre, Saint Petersburg, Russia*
[c] *Tomsk State University, Tomsk, 36 Lenin Avenue, Tomsk, Russian Federation*
[*]Corresponding Author: Sergei I. Tverdokhlebov tverd@tpu.ru

**Abstract**

The aim of this study was to characterize different calcium phosphate coatings and evaluate *in vitro* cell response of these materials to ceramics implants. The physical and chemical properties of calcium phosphate coatings formed by RF-magnetron sputtering of calcium phosphate tribasic, hydroxyapatite, calcium phosphate monobasic, calcium phosphate dibasic dihydrate and calcium pyrophosphate powders were characterized. Cell adhesion and cell viability were examined on calcium phosphate coatings using mesenchymal stem cells. The results of cytotoxicity measurements of the calcium phosphate coatings revealed that only the coating obtained by RF-magnetron sputtering of the calcium phosphate dibasic dihydrate and calcium phosphate tribasic powders possessed lower cell viability than the zirconia substrate. The coating formed by sputtering of the calcium phosphate tribasic powder demonstrated more cells adhered onto its surface compared with other calcium phosphate coatings.

**Key words:** zirconia ceramics, RF-magnetron sputtering, calcium phosphate coatings, cell viability

## 1. Introduction

In recent years, there has been an active interest in improved surgical techniques. There is a shift from using amputation to treatment of patients with endoprosthetics. In the modern world, the number of operated patients with implanted artificial devices into bone or joint tissues increases [1,2]. It allows people to lead a full-fledged life without feeling restricted in movement.

Ceramics in medicine is used to make prosthesis or their individual elements; in medical equipment; for the manufacture of medical instruments, membranes for separation and purification of biological fluids; porous elements for dosed administration of drugs [3]. Ceramics practically does not conduct electric current. It is a material that is electrolytically passive and biologically inert. Ceramics has a high ability of integration with bone tissue, which is advantage over the metal implants [4,5].

Along with the advantages mentioned above, ceramics also has significant disadvantages [3]. Thus, the passivity of ceramics with respect to the living body leads to the fact that a bone may not grow into the implant, and the contact site will be filled with a fibrous tissue that covers the foreign body. High strength of ceramics results in its increased rigidity. The latter leads to load redistribution on the bone in the area of contact of the implant with the bone, as a result of which its destruction occurs. Besides, a ceramic implant itself, under certain conditions, because of the brittleness of ceramics, can be a probable fracture region. Despite these disadvantages, there are areas of traumatology and orthopedic, where ceramic implants have no alternative. First of all, this refers to small joints and teeth prosthesis. «Ceramic steel» made from alumina ($Al_2O_3$) or from zirconia ($ZrO_2$) is the most widely used material.

As yttria decreases the driving force of the t-m transformation, biomedical grade zirconia is usually stabilized with 3 mol% yttria ($Y_2O_3$) (hence 3Y-TZP). Zirconia ceramics stabilized with yttrium oxide has several advantages over other ceramics due to its excellent mechanical and tribological properties together with biocompatibility. Therefore, it finds a wide application as a material for ball heads during total hip joint replacement. Moreover, aesthetically, zirconia ceramics is a more desirable material for dental implants, as compared to gray titanium implants [6–9].

The implant-bone system functioning becomes possible only if the bone tissue is fully integrated with the implant. To improve the integration of the bone tissue with the implant, various bioactive coatings are deposited on the surface of the implant [10,11]. For example, oxidizing treatment [12,13], electrolytic deposition of the layers of calcium-phosphate coatings [14], plasma spraying [15], vacuum ion-plasma deposition [16,17] are effective methods for producing bioinert and bioactive coatings on the surface of metal implants. There are few articles devoted to the modification of ceramic implants. Besides, it is very difficult to choose an optimal coating for a ceramic implant, and also to make an assessment of its clinical effectiveness because of the authors describe the coatings formed by various technologies. The coatings obtained with the help of various technologies differ substantially in physical, chemical, and, consequently, medical and biological properties. We are unware the publications with comparative analysis of coatings of different compositions deposited by the same method on the surface of ceramic substrates. Therefore, the perspectives of the modification of ceramic implants have not been well studied.

Calcium phosphate materials such as hydroxyapatite, tricalcium phosphate (α, β), dicalcium phosphate dihydrate (DCPD), octacalcium phosphate (OCP), etc. possess a great potential as materials for bioactive coatings [11,18,19].

This paper is aimed to provide a comparative study of the coatings deposited by RF-magnetron sputtering of different calcium phosphate powders on zirconia substrates.

Unlike other methods, RF-magnetron sputtering allows deposition of bioactive coatings not only on the surface of metal implants, but also ceramic and polymer substrates [20–22]. Such coatings are characterized by high adhesion to various types of substrates, high elasticity, spatial uniformity, as well as the possibility of obtaining coatings on complex shape implants. Generally, the coatings formed by the plasma-chemical technique are amorphous and can be irreversibly transformed into the crystalline state [23]. This ability can be used to control the biomedical properties of such coatings.

## 2. Materials and methods
*2.1. Materials*

Five different commercially available calcium phosphate powders were used in this study. There was calcium phosphate tribasic ($H_2Ca_{10}O_{26}P_6$), hydroxyapatite ($Ca_{10}(PO_4)_6(OH)_2$), calcium phosphate monobasic ($H_4CaO_8P_2$), calcium phosphate dibasic dihydrate ($HCaO_4P \cdot 2H_2O$) and calcium pyrophosphate ($Ca_2O_7P_2$). All powders were purchased from Sigma-Aldrich Chemie Gmbh (Germany).

The substrates used were 10 mm diameter and 2 mm thick partially yttrium-stabilized zirconium dioxide ($ZrO_2+3$ mol.%$Y_2O_3$) plates. To measure thickness, the Ca-P coatings were sputtered onto Si plates.

Cell adhesion and cytotoxicity of the calcium phosphate coatings were examined using mesenchymal stem cells (MSCs). Fat-derived MSCs were collected from healthy donors immunophenotyped with flow cytometer GuavaEasyCyte6 (Millipore, USA) using CD19, CD34, CD45, CD73, CD90 and CD105 monoclonal antibodies (BD, USA) as previously described [24]. The study was performed according to the Helsinki declaration and approval was obtained from the local Ethics Committees in the Almazov Federal Medical Research Centre. Written consent was obtained from all subjects prior fat tissue biopsy. The cells were maintained in the alpha-MEM medium (PanEco, Russia) supplemented with 10% fetal calf serum (Hyclone, USA), 50 units/mL penicillin and 50 μg/mL streptomycin (Invitrogen, USA) at 37°C and 5% $CO_2$.

*2.2. Methods*
*2.2.1. Coating deposition*

Coating deposition was carried out using the universal magnetron sputtering system based on the set up «Cathod 1M» (assembled in the Hybrid Materials Laboratory of Tomsk Polytechnic University [25], Tomsk, Russia) at a generator frequency of 13.56 MHz. The operational parameters employed during RF magnetron sputter deposition are presented in Table 1. Prior to coating deposition, the ceramics substrates were prepared by grinding and subsequent polishing. After the grinding and polishing steps the disks were ultrasonically cleaned in an ultrasonic bath (Sapfir 5, Russia), and were step-by-step soaked in chloroform and in ethyl alcohol.

*2.2.2. Coating investigations*

*AFM*

Surface morphologies of the coatings were examined by atomic force microscopy (AFM) (Solver-HV, NT-MDT, Russia) operating in the tapping mode. The value of root-mean-square ($R_{ms}$) surface roughness was evaluated over the area of 4 μm.

Table 1. Operational parameters employed during RF magnetron sputter deposition.

| Parameter | Calcium phosphate tribasic $H_2Ca_{10}O_{26}P_6$ | Hydroxyapatite $Ca_{10}(PO_4)_6(OH)_2$ | Calcium phosphate monobasic $H_4CaO_8P_2$ | Calcium phosphate dibasic dehydrate $HCaO_4P \cdot 2H_2O$ | Calcium pyrophosphate $Ca_2O_7P_2$ |
|---|---|---|---|---|---|
| Base pressure (Pa) | 0.007 | 0.007 | 0.007 | 0.007 | 0.007 |
| Working gas | Own outgassing | Argon (99.999%) | Own outgassing | Own outgassing | Own outgassing |
| Power (W) | 1500 | 1500 | 1000 | 1100 | 1500 |
| RF power density (W/cm$^2$) | 6.7 | 6.7 | 4.5 | 4.9 | 6.7 |
| Chamber pressure (Pa) | 0.3-0.5 | 0.3-0.5 | 0.3-0.5 | 0.3-0.5 | 0.3-0.5 |
| Throw distance (mm) | 38 | 38 | 38 | 38 | 38 |
| Deposition time (h) | 7 | 7 | 14 | 7 | 7 |
| Coating thickness (nm) | 100 | 150 | 100 | 80 | 100 |

*Mechanical properties*

The hardness and the elastic modulus of the Ca-P coatings were measured by a nanoindentation test on a NanoTest 600 apparatus (Wrexham, UK) using the Berkovich pyramid.

*Thickness of the Ca-P coatings*

The thickness of the Ca-P coatings was determined by the profilometric method by the profilometer-profilograph Talysurf 5 (Taylor-Hobson, UK).

*XRD analysis*

The crystal structure of the samples was investigated using X-ray diffraction analysis with the Shimadzu XRD 6000 diffractometer. The samples were exposed to the monochromatic Cu K-alpha (1.54056 Å) radiation. The accelerating voltage and the beam current were set to 40 kV and 30 mA, respectively. The scanning angle range, scanning step size and the signal collection time were 6–55, 0.0200 and 1.5 s respectively.

*Elemental analysis*

Elemental composition studies were performed with the scanning electron microscope Quanta 200 3D (FEI Company, Hillsboro, USA) integrated with an energy dispersive X-ray (EDX) detector (JSM-5900LV; JEOL Ltd, Tokyo, Japan) in low vacuum with an electron beam accelerating voltage of 10 kV.

*FTIR analysis*

The chemical structure of the samples was studied using the Attenuated total reflectance Fourier transform infrared (FTIR) spectroscopy using the Nicolet 6700 system (Thermo Scientific, USA) in the range of 800-2000 $cm^{-1}$ with a resolution of 1 $cm^{-1}$.

*Wettability of the Ca-P coatings*

Wettability of the Ca-P coatings was studied with the "Easy Drop" device (Krüss, Germany) by the method of the "sit" drop by measuring the contact angle of a liquid drop with volume of 3 µl placed onto the investigated surface. Measurements of the wetting boundary angle (the contact angle) were carried out one minute after placing the liquid on the surface. In order to avoid contamination of the surface and distortion of measurements results, measurements of the contact angle were carried out immediately after the surface modification. Dymethyl formamide and water were used as the wetting liquids. The total surface energy, it's polar and dispersion components were calculated by the Owens-Wendt-Rabel-Kaelble (OWRK) method.

*Cell viability*

Zirconia samples were placed into the 24 well culture plates where MSCs were added in the culture medium in an amount of 60 000 cells / well. The cultivation was carried out under incubator conditions, in a humid environment at 37 °C and the $CO_2$ content of 5% for 72 hours. Subsequently, samples were treated with accutase for subsequent cell viability analysis by laser flow cytometry. Cell suspensions were stained with Annexin V FITC (Biolegend) and Propidium Iodid (Sigma Aldrich) according to the manufacturer's recommendations. Flow cytometry data analysis of the samples was performed on a GuavaEasyCyte8 flow cytometer. Analysis of the obtained results was made using the Kaluza software (Beckman Coulter). Double-positive events were assessed as cells in the state of late apoptosis or necrosis, positive events according to Annexin V as cells in the state of early apoptosis and double-negative events as living cells.

*Fluorescence microscopy of the cells*

To analyse the effect of surface modification on cell adhesion and attachment, focal adhesion and cytoskeletal proteins, vinculin and α-actin were studied using fluorescence microscopy. Samples with cells after 72 hours of incubation were discarded from the medium; samples were washed with phosphate buffer solution (PBS) and fixed with 4% paraformaldehyde for 20 min. Cells were permeabilised using triton ×100. The cells were rinsed with PBS, blocked with 10% goat serum in PBS for 30 min at room temperature and incubated with the anti-vinculin antibody (Thermo Scientific) at 1:200 dilution for 2 hours. After 3 PBS washes the cells were incubated with Alexa Fluor 488 goat anti-mouse IgG (H+L) (Invitrigen) at 1:200 dilution for 1 h at room temperature in the dark. For α-actin MSCs were stained with α-SMA (Diagnostic BioSystems) and with Alexa Fluor 546 goat anti-mouse IgG (H+L) (Invitrigen). The cells were washed three times in PBS (5 min each) and stained with 4′,6 diamidino-2-phenylindole (DAPI) for nuclear visualization. Finally, the cells were washed and viewed using the CarlZeiss Axio Observer fluorescence microscope. The images were collected and processed with Zen Software. Morphometric analysis was performed using Image J Software.

*Statistical analysis*

Statistical analysis was performed using STATISTICA 7.0 software (Statsoft, Tulsa, USA). The data are presented as mean ± standard deviation. The significance of difference was calculated using the oneway ANOVA test and the Mann-Whitney U-test.

## 3. Results and discussion

The data obtained by atomic force microscopy indicates that the zirconia substrate is characterized by a relatively equiaxed grains with an average size of 130 nm and the $R_{ms}$ of 30 nm (Fig. 1a). Modification of the zirconia substrate by means of sputtering of the calcium phosphate powders leads to a change in its initial morphology. The surface of the coatings formed by the deposition of calcium phosphate tribasic powder is wavy owing to the alternated hillocks and valleys (Fig. 1b). At higher magnification, the structural elements in the form of thin plates (a plate-like structure) with the length of 80 nm and the width of 30 nm become clearly visible. The $R_{ms}$ of such coatings is 7.00 nm. The surface morphology of the remaining calcium phosphate coatings qualitatively different from the previous one is a globular formation of a differing size. Thus, the coating surface formed by sputtering of hydroxyapatite powder consists of spherical globules with a mean size of ~ 20 nm. The $R_{ms}$ of coating is 9.10 nm (Fig. 1c). The calcium-phosphate coatings deposited by sputtering of calcium phosphate monobasic, calcium phosphate dibasic dihydrate and calcium pyrophosphate powders have a more homogeneous structure with an average element size of 13.50, 10.60 and 7.60 nm, respectively (Fig. 1d,e,f). The roughness of these coatings with respect to the $R_{ms}$ parameter practically does not differ and is 3.60, 3.80 and 3.60 nm, respectively.

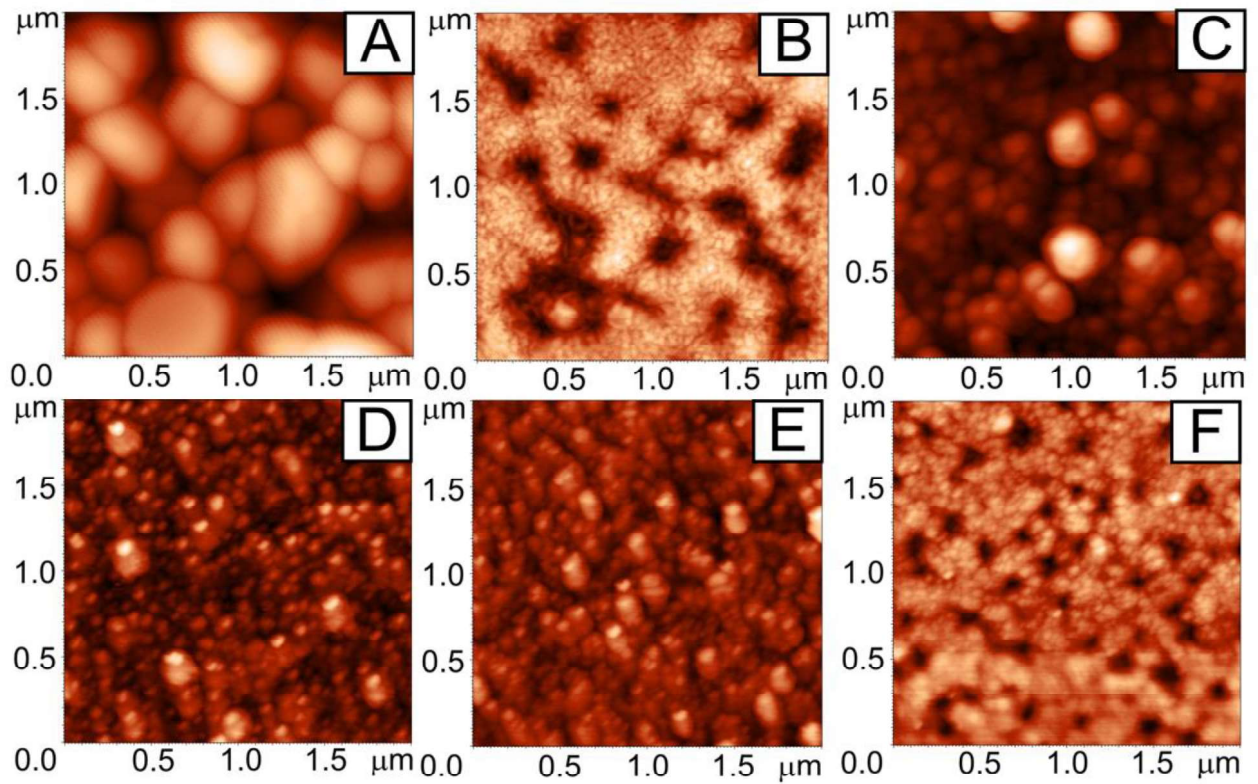

Fig.1 AFM images of the CaP coatings on the ZrO2 substrates.

The mechanical properties of the investigated calcium-phosphate coatings are presented in Table 2. To exclude the influence of the substrate material during the measurement of the mechanical properties of the coatings, the load on the indenter did not exceed 0.5 mN. Analysis of the nanoindentation test data showed that the zirconia substrate and the coating obtained by sputtering of the calcium phosphate tribasic powder had the largest mechanical properties among the investigated samples. In turn, the nanohardness value and the modulus of elasticity of the coatings deposited by sputtering of hydroxyapatite, calcium phosphate dibasic dehydrate and calcium pyrophosphate powders were practically the same. The coating formed from calcium phosphate monobasic powder had the lowest values of nanohardness and the modulus of elasticity among all the samples.

The high elastic properties of the zirconia substrate and the coating deposited by sputtering of the calcium phosphate tribasic powder are confirmed by a higher (as compared to other calcium-phosphate coatings) elastic recovery value ($R$). The latter depends on the modulus of elasticity of the material. In non-crystalline materials, the elastic component arises due to the elastic displacement of the points of the structural grid under an external load. Its value is proportional to the mechanical compliance of the material.

Based on the literature data, the modulus of elasticity of the human trabecular and cortical bone varies between 15-19 GPa and 13-26 GPa, respectively [26–28]. In turn, the hardness value for the human cortical bone, according to different data, is 0.30-0.80 GPa [29]. The comparison the data of other authors with the data obtained for the calcium-phosphate coatings investigated in this article is not possible because of difference in the measurement range. According to the international standard ISO 14577, the nanoindentation method allows where $P_{max}$ – the maximum load applied to the indenter, $h_{max}$ – the maximum penetration depth of the indenter into the coating, $H$ – nanohardness, $E^*$ – the modulus of elasticity, $R$ – the print elastic recovery value.

Table 2. The mechanical properties of calcium phosphate coatings.

| | $P_{max}$, mN | $h_{max}$, nm | $H$, GPa | $E^*$, GPa | $R$ |
|---|---|---|---|---|---|
| Calcium phosphate tribasic | | 38 | 7.33±2.59 | 115±25 | 0.44 |
| Hydroxyapatite | 0.5 | 59 | 3.44±0.4 | 78±10 | 0.25 |
| Calcium phosphate monobasic | | 91 | 1.67±0.31 | 68±9 | 0.12 |
| Calcium phosphate dibasic dihydrate | | 63 | 3.10±1.19 | 99±26 | 0.17 |
| Calcium pyrophosphate | | 64 | 3.15±1.13 | 83±23 | 0.21 |
| ZrO$_2$ substrate | | 34 | 8.88±2.13 | 127±23 | 0.54 |

measurement of the mechanical characteristics in three ranges: nano range: hmax ≤ 200 nm; micro range: $h_{max}$ > 200 nm and $F_{max}$ < 2N and macro range: 2 N ≤ $F_{max}$ ≤ 30 kN [30]. In this paper, the mechanical properties of calcium-phosphate coatings were investigated at a maximum load of 0.5 mN and a maximum indenter penetration depth of 90 nm, which were the optimal parameters for thin coatings. In the literature on the mechanical characteristics of bone tissue, as a rule, bulk samples are examined at a microscale.

In addition to the comparison of the mechanical properties of calcium-phosphate coatings with the bone tissue, it is necessary to take into account the difference in the mechanical properties between the coating and the substrate. The significant difference in the mechanical characteristics of the substrate material and the coating can be responsible for the stresses occurring at the interface «coating-substrate», which can result in the coating delamination [31,32].

XRD analysis showed all the calcium phosphate coatings to be amorphous.

The elemental composition of calcium-phosphate coatings is presented in Table 3. Special attention should be paid to the content of the elements such as Ca and P, which play an important role in the processes of cell activity [32]. Even more, these elements contribute to the formation of the bone tissue [33]. Thus, the greatest amount of Ca is observed in the coating deposited by sputtering of the hydroxyapatite powder. A slightly less amount of Ca is found in the coating formed from of calcium phosphate dibasic dihydrate powder. The other calcium phosphate coatings have approximately the same content of Ca (10-14 at.%).

A similar tendency is observed with the content of P calcium phosphate coatings. The largest amount of this element is contained in the composition of the coatings obtained by sputtering of hydroxyapatite and calcium phosphate dibasic dihydrate powders. The presence of such elements as Y, Zr and O is due to the substrate composition. The presence of C in the composition of calcium-phosphate coatings is associated with the process of their deposition.

Using IR spectroscopy enabled revealing weak absorption bands at 560 and 600 cm$^{-1}$ along with an absorption band between 1000-1200 cm$^{-1}$ in spectra of all investigated coatings (Fig. 2). The intensity of the latter is maximal for the coating deposited by sputtering of the calcium phosphate tribasic powder and gradually decreases for the other calcium phosphate coatings. The bands mentioned above correspond to the valence vibrations of the $PO_4^{3-}$ group [34,35]. There were no other characteristic absorption bands in IR spectra of the calcium-phosphate coatings.

Table 3. The elemental composition of calcium phosphate coatings.

| Calcium phosphate coatings | C, at.% | O, at.% | Y, at.% | P, at.% | Zr, at.% | Ca, at.% |
|---|---|---|---|---|---|---|
| Calcium phosphate tribasic | 17.17 | 31.95 | 1.37 | 5.97 | 29.65 | 13.79 |
| Hydroxyapatite | 12.63 | 31.42 | 1.04 | 11.11 | 17.21 | 26.58 |
| Calcium phosphate monobasic | 19.52 | 30.77 | 1.65 | 5.87 | 31.50 | 10.69 |
| Calcium phosphate dibasic dihydrate | 15.75 | 30.58 | 1.39 | 8.25 | 27.95 | 16.08 |
| Calcium pyrophosphate | 14.81 | 32.44 | 1.60 | 6.69 | 32.26 | 12.20 |
| $ZrO_2$ substrate | 19.62 | 27.71 | 4.95 | - | 47.71 | - |

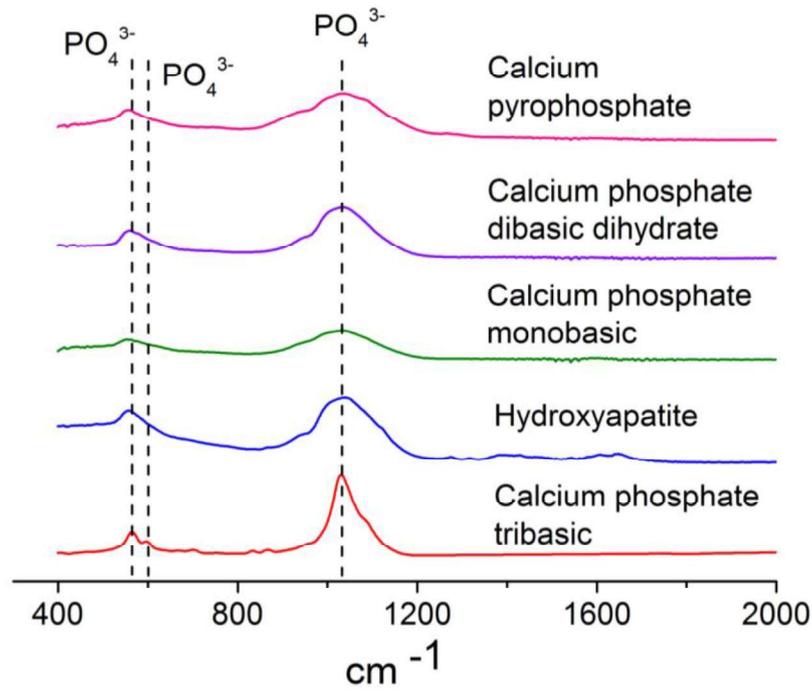

Fig.2 The FTIR spectra of the as-deposited calcium phosphate coatings.

The deposition of calcium-phosphate coatings on the zirconia substrate leads to a change in its free surface energy (Table 4). The analysis of the data presented in Table 2 showed that the coating deposited from the hydroxyapatite powder possess the largest value of the free surface energy and the smallest one possessed the coating obtained by sputtering of calcium pyrophosphate powder. This fact is in good agreement with the roughness data: the greatest value of the roughness corresponds to the largest value of free surface and vice versa. It is well-known that the microrelief and the surface roughness are closely related to the value of the free surface energy [36] and, consequently, to the surface wettability [37,38]. The roughness of the surface causes an increase in the area of the actual contact of the liquid with the solid as compared to the smoother surface. In turn, an increase in the area of the actual contact results in a proportional increase in the specific free surface energy of the rough surface compared to a smoother surface.

Table 4. The contact angles and surface energy of calcium phosphate coatings.

| Calcium phosphate coatings | Contact angle of water Θ, degrees | Contact angle of dymethyl formamide Θ, degrees | Free surface energy γ, mJ/m$^2$ | Dispersion component γ$^d$, mJ/m$^2$ | Polar component γ$^p$, mJ/m$^2$ |
|---|---|---|---|---|---|
| Calcium phosphate tribasic | 98.0±1.25 | 31.7±4.31 | 38.58±1.48 | 38.47±1.43 | 0.11±0.05 |
| Hydroxyapatite | 25.3±6.16 | 7.8±2.93 | 70.62±2.41 | 6.98±0.51 | 63.65±1.90 |
| Calcium phosphate monobasic | 76.1±3.89 | 39.8±6.66 | 30.45±2.49 | 17.65±1.52 | 12.79±0.96 |
| Calcium phosphate dibasic dihydrate | 103.0±4.74 | 56.7±3.99 | 25.11±1.29 | 24.62±1.14 | 0.50±0.15 |
| Calcium pyrophosphate | 95.7±3.75 | 57.2±2.20 | 22.35±0.86 | 19.45±0.60 | 2.90±0.26 |
| ZrO$_2$ substrate | 98.4±2.98 | 53.8±2.44 | 25.01±0.84 | 23.64±0.69 | 1.37±0.16 |

Thus, the authors [36] observed a linear increase in the free surface energy of the WNx films with increasing surface roughness ($R_{ms}$). In [39], it was shown that the substrate roughness affected the superhydrophobic behavior of polytetrafluoroethylene thin films. The authors demonstrated that the large contact angle of the water and the low hysteresis of the contact angle corresponded to the low values of the free surface energy. Moreover, the authors established an optimum $R_{ms}$ parameter, at which the polytetrafluoroethylene films had superhydrophobic properties. This demonstrates that the data obtained in our study of calcium phosphate coatings are in good agreement with the data obtained by other authors for other coatings.

Wettability studies (water data) of the calcium-phosphate coatings showed that the coatings deposited from the powders of calcium phosphate dibasic dehydrate, calcium phosphate tribasic and calcium pyrophosphate tended to have hydrophobic properties. And on the contrary, the coatings formed using the hydroxyapatite and calcium phosphate monobasic powders possess the hydrophylic properties. For these coatings, the polar component of the free surface energy is maximal, which indicates a high wettability of these coatings by polar liquids, for example, such as blood.

Free surface energy and surface wettability play an important role in the interaction of the implant surface with the biological environment [39–41]. In the literature there are a large number of studies devoted to that fact that a high free surface energy or good wettability (hydrophilicity) contributes to improved cell adhesion. Conversely, the surfaces with a low free surface energy and hydrophobic properties are unfavorable for attachment of cells and their further spreading over the surface [42–44]. Some researchers have found that there is an optimal contact angle of the order of 60-70° at which the cells preferably adhere to surface and, moreover, the directed proliferation takes place [45,46]. However, the above data are not always confirmed [43,45,47,48].

The number of the MSCs adhered on the calcium phosphate coatings as well as overall cell area, single cell area are presented in Table 5. Only the samples formed by using the Calcium phosphate tribasic powder showed a significantly higher cell adhesion rate which was performed by higher number of cells as well as overall area covered by cells. The other calcium-phosphate coatings demonstrated a similar cell adhesion.

Fluorescence analysis of the cell morphologies showed that the cells adhered on the substrate were spread over the material and formed a continuous cell layer on the surface (Fig. 3a). By staining vinculin except for a diffuse staining of the cytoplasm, the numerous focal adhesions

Table 5. Number of the mesenchymal stem cells adhered on the calcium phosphate coatings, cells/mm$^2$, (Mean±SD).

| Calcium phosphate coatings | Cells/mm$^2$ | Overall cell area, % of total surface | Single cell area, μm$^2$ |
|---|---|---|---|
| Calcium phosphate tribasic | 284.33±84.53* | 77.73±8.64* | 897±163 |
| Hydroxyapatite | 143.33±66.66 | 53.91±21.01 | 1223±214 |
| Calcium phosphate monobasic | 210.00±105.22 | 61.18±17.36 | 1021±279 |
| Calcium phosphate dibasic dihydrate | 190.25±45.55 | 58.05±12.31 | 969±114 |
| Calcium pyrophosphate | 212.18±69.39 | 64.55±12.34 | 1015±174 |
| ZrO$_2$ substrate | 185.25±79.68 | 54.40±25.20 | 970±403 |

*$p<0.05$ comparing to ZrO$_2$ substrate

Table 6. The results of cell viability on the surface of calcium phosphate coatings, (Mean±SD)

| Calcium phosphate coatings | Viable, % | Early apoptosis, % | Late apoptosis/Necrosis, % |
|---|---|---|---|
| Calcium phosphate tribasic | 59.40±3.11* | 2.20±0.62 | 35.54±4.11* |
| Hydroxyapatite | 74.01±1.36 | 3.37±2.44 | 20.44±3.62 |
| Calcium phosphate monobasic | 66.79±9.96 | 2.60±1.76 | 28.85±8.07 |
| Calcium phosphate dibasic dihydrate | 55.49±11.39* | 3.49±1.33* | 38.97±9.74* |
| Calcium pyrophosphate | 71.49±9.39 | 1.84±0.69 | 25.43±8.64 |
| $ZrO_2$ substrate | 77.32±4.23 | 1.45±0.58 | 20.36±3.34 |

*p=0.05 comparing to $ZrO_2$ substrate

between the cells and with the material were performed. Actin in the trabecular meshwork was observed in almost all cells (Fig. 3). The vinculin was predominantly present in the cytoplasm although the focal adhesion was not clearly seen.

Comparison of the data on cell adhesion, morphology with the value of their free surface energy and the wettability does not allow making an unambiguous conclusion about the presence of any relation between these quantities. The latter is explained by the fact that the whole complex of the physical and chemical characteristics of calcium-phosphate coatings affects the value of cell adhesion.

The cytotoxicity tests of the calcium phosphate coatings revealed that the largest number of viable cells had been removed from the surface of the zirconia substrate (Table 6). The cytotoxicity of the coatings obtained by sputtering of the calcium pyrophosphate, hydroxyapatite and calcium phosphate monobasic powders did not differ significantly from the one for the substrate. At the same time, the samples with coatings formed by sputtering of the calcium phosphate dibasic dihydrate and calcium phosphate tribasic powders showed a significantly smaller percentage of viable cells and a larger number of the late apoptotic/necrotic cells.

One of the main reason for cell death is the ROS-mediated-oxidative stress [49,50]. So, in the process of sputtering of the calcium phosphate dibasic dehydrate powder, the generation of $H^+$ and $OH^-$ free radicals is possible due to the dissociation of the water molecules which are incorporated into the powder. Free radicals contribute to lipid peroxidation of the membranes and / or cross-linking of the proteins, which results in cell death. However, the above assumption has not been supported by the IR spectroscopy data. Analysis of the IR spectrum of the coating formed from the calcium phosphate dibasic dehydrate powder did not discover absorption bands corresponding to $OH^-$ and $H_2O$. The latter can be attributed to a small amount of $H_2O$.

Another likely cause of the cell death when they interact with calcium phosphate coatings can be the release of $Ca^{2+}$ and $PO_4^{3-}$ ions during incubation. The capture of these ions by the cell leads to a change in the permeability of the mitochondrial membrane. The latter, in turn, can result in cell death. The dependence of the cell viability on the level of the released $Ca^{2+}$ and $PO_4^{3-}$ ions has been previously shown in [51–53]. The authors [51] especially emphasize that an increase in cell death occurs only when the concentration of both $Ca^{2+}$ and $PO_4^{3-}$ ions rises. The change in the concentration of only one type of ions in the extracellular space does not lead to a similar effect.

However, there is no evidence for this assumption. In the IR spectra of the investigated calcium-phosphate coatings, only the peaks of $PO_4^{3-}$ ions are present, whereas $Ca^{2+}$ ions were not detected. Taking into account the intensity of these peaks, it can be concluded that the greatest number of the phosphate ions is incorporated into coatings deposited by sputtering of the powders of calcium phosphate tribasic, hydroxyapatite and calcium phosphate dibasic dihydrate.

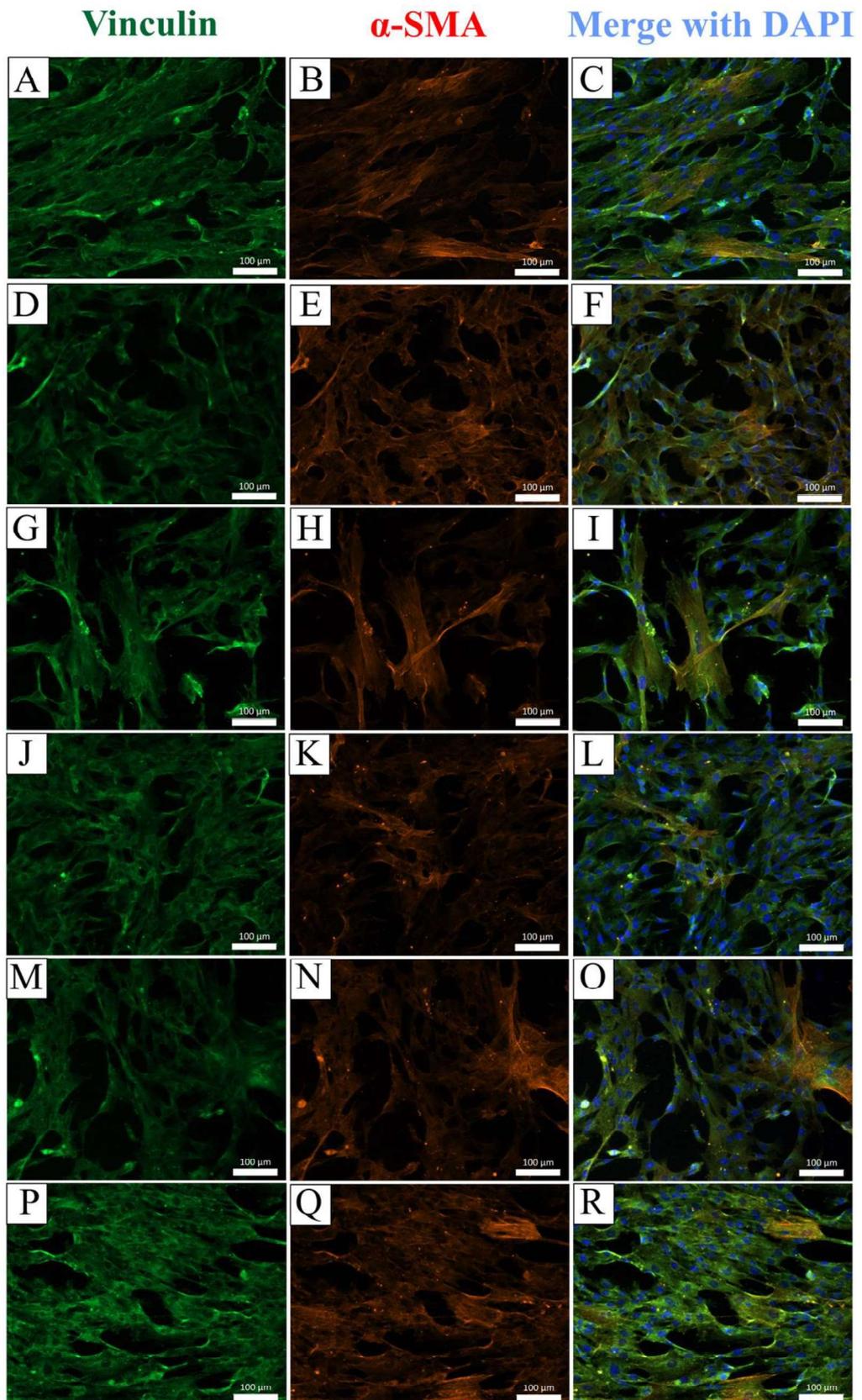

Fig.3. Immunofluorescence staining of the mesenchymal stem cells (MSCs) cultured on the ZrO$_2$ substrates with different Ca-P coatings. A,B,C – ZrO$_2$; D,E,F - Calcium phosphate tribasic; G,H,I – Hydroxyapatite, G,K,L- Calcium phosphate monobasic; M,N,O - Calcium phosphate dibasic dehydrate; P,Q,R - Calcium pyrophosphate. Scale bar: 100 μm.

Less pronounced peaks are registered in the spectrum of the coating obtained from the calcium pyrophosphate powder. Finally, the least apparent peak of the phosphate ions is present in the IR spectrum of the coating formed from the calcium phosphate monobasic powder. In accordance with the IR spectroscopy data, we can say nothing about the amount of $Ca^{2+}$ ions. Moreover, the analysis of the elemental composition of the coatings, on the contrary, indicates a large content of the Ca and P elements in the coatings, which show the cytotoxicity similar to the zirconia substrate. The facts mentioned above do not allow one to estimate of the amount of released $PO_4^{3-}$ and $Ca^{2+}$ ions from the coatings into the surrounding intercellular space during the incubation. Therefore, additional studies are needed to reveal the cytotoxicity mechanism of the calcium-phosphate coatings when they interact with the cells.

## 4. Conclusions

A comparative study of calcium phosphate coatings deposited on the zirconia substrate by RF-magnetron sputtering of calcium phosphate tribasic, hydroxyapatite, calcium phosphate monobasic, calcium phosphate dibasic dihydrate and calcium pyrophosphate powders was carried out. There was no correlation between the physico-chemical properties of the coatings and the biological response of cells in this work.

It is shown that the cytotoxicity of zirconia substrate and calcium phosphate coatings formed from hydroxyapatite, calcium pyrophosphate and calcium phosphate monobasic powders were not significantly different. A significantly smaller percentage of viable cells were observed on the surface of the coatings obtained by sputtering of the calcium phosphate dibasic dihydrate and calcium phosphate tribasic powders. Meanwhile higher number of adhered cells as well as overall cell area were detected on the surface of the coatings sputtered by the calcium phosphate tribasic powder.

Thus, in this work the first stage of *in vitro* evaluation of the coatings deposited on the zirconia substrate surfaces by RF-magnetron sputtering of different calcium-phosphate powders was completed. To make a judgement about the advisability of applying calcium-phosphate coatings to ceramic implants and to assess their clinical effectiveness, it is necessary to investigate the coatings *in vivo* on the ability to stimulate bone formation processes.

**Acknowledgements**

The research is carried out at Tomsk Polytechnic University within the framework of Tomsk Polytechnic University Competitiveness Enhancement Program grant and project VIU-316/2017.

**References:**

[1]  S. Bauer, P. Schmuki, K. von der Mark, J. Park, Engineering biocompatible implant surfaces, Prog. Mater. Sci. 58 (2013) 261–326. doi:10.1016/j.pmatsci.2012.09.001.

[2]  N. Hou, H. Perinpanayagam, M. Mozumder, J. Zhu, Novel Development of Biocompatible Coatings for Bone Implants, Coatings. 5 (2015) 737–757. doi:10.3390/coatings5040737.

[3]  L. Treccani, T. Yvonne Klein, F. Meder, K. Pardun, K. Rezwan, Functionalized ceramics for biomedical, biotechnological and environmental applications, Acta Biomater. 9 (2013) 7115–7150. doi:10.1016/j.actbio.2013.03.036.

[4]  A. Scarano, F. Di Carlo, M. Quaranta, A. Piattelli, Bone Response to Zirconia Ceramic Implants: An Experimental Study in Rabbits, J. Oral Implantol. 29 (2003) 8–12. doi:10.1563/1548-1336(2003)029<0008:BRTZCI>2.3.CO;2.

[5]  M. Bächle, F. Butz, U. Hübner, E. Bakalinis, R.J. Kohal, Behavior of CAL72 osteoblast-like cells cultured on zirconia ceramics with different surface topographies, Clin. Oral Implants Res. 18 (2007) 53–59. doi:10.1111/j.1600-0501.2006.01292.x.

[6] R.J. Kohal, W. Att, M. Bächle, F. Butz, Ceramic abutments and ceramic oral implants. An update, Periodontol. 2000. 47 (2008) 224–243. doi:10.1111/j.1600-0757.2007.00243.x.

[7] H. Ananth, V. Kundapur, H.S. Mohammed, M. Anand, G.S. Amarnath, S. Mankar, A review on biomaterials in dental implantology, Int. J. Biomed. Sci. 11 (2015) 113–120.

[8] P.F. Manicone, P. Rossi Iommetti, L. Raffaelli, An overview of zirconia ceramics: Basic properties and clinical applications, J. Dent. 35 (2007) 819–826. doi:10.1016/j.jdent.2007.07.008.

[9] E.E. Daou, The Zirconia Ceramic: Strengths and Weaknesses, Open Dent. J. 8 (2014) 33–42. doi:10.2174/1874210601408010033.

[10] T. Hanawa, Biofunctionalization of titanium for dental implant, Jpn. Dent. Sci. Rev. 46 (2010) 93–101. doi:10.1016/j.jdsr.2009.11.001.

[11] B. Zhang, D. Myers, G. Wallace, M. Brandt, P. Choong, Bioactive Coatings for Orthopaedic Implants—Recent Trends in Development of Implant Coatings, Int. J. Mol. Sci. 15 (2014) 11878–11921. doi:10.3390/ijms150711878.

[12] B.C. de V. Gurgel, P.F. Gonçalves, S.P. Pimentel, F.H. Nociti, E.A. Sallum, A.W. Sallum, M.Z. Casati, An Oxidized Implant Surface May Improve Bone-to-Implant Contact in Pristine Bone and Bone Defects Treated With Guided Bone Regeneration: An Experimental Study in Dogs, J. Periodontol. 79 (2008) 1225–1231. doi:10.1902/jop.2008.070529.

[13] G. Wang, J. Li, K. Lv, W. Zhang, X. Ding, G. Yang, X. Liu, X. Jiang, Surface thermal oxidation on titanium implants to enhance osteogenic activity and in vivo osseointegration, Sci. Rep. 6 (2016) 31769. doi:10.1038/srep31769.

[14] M.A. Lopez-Heredia, P. Weiss, P. Layrolle, An electrodeposition method of calcium phosphate coatings on titanium alloy, J. Mater. Sci. Mater. Med. 18 (2007) 381–390. doi:10.1007/s10856-006-0703-8.

[15] C.P.A.T. Klein, J.G.C. Wolke, J.M.A. De Blieck-Hogervorst, K. de Groot, Calcium phosphate plasma-sprayed coatings and their stability: An in vivo study, J. Biomed. Mater. Res. 28 (1994) 909–917. doi:10.1002/jbm.820280810.

[16] T. Lu, Y. Qiao, X. Liu, Surface modification of biomaterials using plasma immersion ion implantation and deposition, Interface Focus. 2 (2012) 325–336. doi:10.1098/rsfs.2012.0003.

[17] A.A. Ilyin, S. V. Skvortsova, L.M. Petrov, Y. V. Chernyshova, E.A. Lukina, Effect of vacuum ion-plasma treatment on the electrochemical corrosion characteristics of titanium-alloy implants, Russ. Metall. 2007 (2007) 423–427. doi:10.1134/S0036029507050138.

[18] S. V. Dorozhkin, M. Epple, Biological and Medical Significance of Calcium Phosphates, Angew. Chemie Int. Ed. 41 (2002) 3130–3146. doi:10.1002/1521-3773(20020902)41:17<3130::AID-ANIE3130>3.0.CO;2-1.

[19] D.O. Costa, B.A. Allo, R. Klassen, J.L. Hutter, S.J. Dixon, A.S. Rizkalla, Control of Surface Topography in Biomimetic Calcium Phosphate Coatings, Langmuir. 28 (2012) 3871–3880. doi:10.1021/la203224a.

[20] S.I. Tverdokhlebov, E.N. Bolbasov, E.V. Shesterikov, A.I. Malchikhina, V.A. Novikov, Y.G. Anissimov, Research of the surface properties of the thermoplastic copolymer of vinilidene fluoride and tetrafluoroethylene modified with radio-frequency magnetron sputtering for medical application, Appl. Surf. Sci. 263 (2012) 187–194. doi:10.1016/j.apsusc.2012.09.025.

[21] J.Z. Shi, C.Z. Chen, H.J. Yu, S.J. Zhang, Application of magnetron sputtering for producing bioactive ceramic coatings on implant materials, Bull. Mater. Sci. 31 (2008) 877–884. doi:10.1007/s12034-008-0140-z.

[22] T. Narushima, K. Ueda, T. Goto, H. Masumoto, T. Katsube, H. Kawamura, C. Ouchi, Y. Iguchi, Preparation of Calcium Phosphate Films by Radiofrequency Magnetron Sputtering, Mater. Trans. 46 (2005) 2246–2252. doi:10.2320/matertrans.46.2246.

[23] K. VANDIJK, H. SCHAEKEN, J. WOLKE, J. JANSEN, Influence of annealing

temperature on RF magnetron sputtered calcium phosphate coatings, Biomaterials. 17 (1996) 405–410. doi:10.1016/0142-9612(96)89656-6.

[24] R.I. Dmitrieva, I.R. Minullina, A.A. Bilibina, O. V. Tarasova, S. V. Anisimov, A.Y. Zaritskey, Bone marrow- and subcutaneous adipose tissue-derived mesenchymal stem cells: Differences and similarities, Cell Cycle. 11 (2012) 377–383. doi:10.4161/cc.11.2.18858.

[25] S.I. Tverdokhlebov, E.N. Bolbasov, E.V. Shesterikov, L.V. Antonova, A.S. Golovkin, V.G. Matveeva, D.G. Petlin, Y.G. Anissimov, Modification of polylactic acid surface using RF plasma discharge with sputter deposition of a hydroxyapatite target for increased biocompatibility, Appl. Surf. Sci. 329 (2015) 32–39. doi:10.1016/j.apsusc.2014.12.127.

[26] C.H. Turner, J. Rho, Y. Takano, T.Y. Tsui, G.M. Pharr, The elastic properties of trabecular and cortical bone tissues are similar: results from two microscopic measurement techniques, J. Biomech. 32 (1999) 437–441. doi:10.1016/S0021-9290(98)00177-8.

[27] J.-Y. Rho, M.E. Roy, T.Y. Tsui, G.M. Pharr, Elastic properties of microstructural components of human bone tissue as measured by nanoindentation, J. Biomed. Mater. Res. 45 (1999) 48–54. doi:10.1002/(SICI)1097-4636(199904)45:1<48::AID-JBM7>3.0.CO;2-5.

[28] Z. Fan, J.G. Swadener, J.Y. Rho, M.E. Roy, G.M. Pharr, Anisotropic properties of human tibial cortical bone as measured by nanoindentation, J. Orthop. Res. 20 (2002) 806–810. doi:10.1016/S0736-0266(01)00186-3.

[29] L. Fricoteaux, Thèse présentée pour l ' obtention du grade de Docteur de l ' UTC, (2012).

[30] M. Kurdej, Thèse présentée pour l ' obtention du grade de Docteur de l ' UTC, (2013). https://tel.archives-ouvertes.fr/tel-01186735.

[31] B. Cofino, P. Fogarassy, P. Millet, A. Lodini, Thermal residual stresses near the interface between plasma-sprayed hydroxyapatite coating and titanium substrate: Finite element analysis and synchrotron radiation measurements, J. Biomed. Mater. Res. 70A (2004) 20–27. doi:10.1002/jbm.a.30044.

[32] J.W. Hutchinson, H.M. Jensen, Delamination of thin films, Engineering. 3 (1996) 45. doi:10.1.1.366.4011.

[33] S. V. Gnedenkov, S.L. Sinebryukhov, A. V. Puz′, V.S. Egorkin, R.E. Kostiv, In vivo study of osteogenerating properties of calcium-phosphate coating on titanium alloy Ti–6Al–4V, Biomed. Mater. Eng. 27 (2017) 551–560. doi:10.3233/BME-161608.

[34] S. Raynaud, E. Champion, D. Bernache-Assollant, P. Thomas, Calcium phosphate apatites with variable Ca/P atomic ratio I. Synthesis, characterisation and thermal stability of powders, Biomaterials. 23 (2002) 1065–1072. doi:10.1016/S0142-9612(01)00218-6.

[35] I. Mobasherpour, M.S. Heshajin, A. Kazemzadeh, M. Zakeri, Synthesis of nanocrystalline hydroxyapatite by using precipitation method, J. Alloys Compd. 430 (2007) 330–333. doi:10.1016/j.jallcom.2006.05.018.

[36] C.-W. Fan, S.-C. Lee, Surface Free Energy Effects in Sputter-Deposited WNx Films, Mater. Trans. 48 (2007) 2449–2453. doi:10.2320/matertrans.MRA2007095.

[37] K.J. Kubiak, M.C.T. Wilson, T.G. Mathia, P. Carval, Wettability versus roughness of engineering surfaces, Wear. 271 (2011) 523–528. doi:10.1016/j.wear.2010.03.029.

[38] A.S.H. Moita, A.L.N. Moreira, A.S. Moita, Influence of Surface Properties on the Dynamic Behavior of Impacting Droplets, (2003).

[39] H.C. Barshilia, D.K. Mohan, N. Selvakumar, K.S. Rajam, Effect of substrate roughness on the apparent surface free energy of sputter deposited superhydrophobic polytetrafluoroethylene thin films, Appl. Phys. Lett. 95 (2009) 33116. doi:10.1063/1.3186079.

[40] G. Zhao, Z. Schwartz, M. Wieland, F. Rupp, J. Geis-Gerstorfer, D.L. Cochran, B.D. Boyan, High surface energy enhances cell response to titanium substrate microstructure, J. Biomed. Mater. Res. Part A. 74A (2005) 49–58. doi:10.1002/jbm.a.30320.

[41] D. Kubies, L. Himmlová, T. Riedel, E. Chánová, K. Balík, M. Douděrová, J. Bártová, V. Pešáková, The interaction of osteoblasts with bone-implant materials: 1. The effect of physicochemical surface properties of implant materials., Physiol. Res. 60 (2011) 95–111. http://www.ncbi.nlm.nih.gov/pubmed/20945966.

[42] X. Liu, J.Y. Lim, H.J. Donahue, R. Dhurjati, A.M. Mastro, E.A. Vogler, Influence of substratum surface chemistry/energy and topography on the human fetal osteoblastic cell line hFOB 1.19: Phenotypic and genotypic responses observed in vitro☆, Biomaterials. 28 (2007) 4535–4550. doi:10.1016/j.biomaterials.2007.06.016.

[43] J.Y. Lim, X. Liu, E.A. Vogler, H.J. Donahue, Systematic variation in osteoblast adhesion and phenotype with substratum surface characteristics, J. Biomed. Mater. Res. 68A (2004) 504–512. doi:10.1002/jbm.a.20087.

[44] P. Parhi, A. Golas, E.A. Vogler, Role of Proteins and Water in the Initial Attachment of Mammalian Cells to Biomedical Surfaces: A Review, J. Adhes. Sci. Technol. 24 (2010) 853–888. doi:10.1163/016942409X12598231567907.

[45] T. Groth, G. Altankov, Studies on cell-biomaterial interaction: role of tyrosine phosphorylation during fibroblast spreading on surfaces varying in wettability, Biomaterials. 17 (1996) 1227–1234. doi:10.1016/0142-9612(96)84943-X.

[46] Y. Tamada, Y. Ikada, Fibroblast growth on polymer surfaces and biosynthesis of collagen, J. Biomed. Mater. Res. 28 (1994) 783–789. doi:10.1002/jbm.820280705.

[47] M. Padial-Molina, P. Galindo-Moreno, J.E. Fernández-Barbero, F. O'Valle, A.B. Jódar-Reyes, J.L. Ortega-Vinuesa, P.J. Ramón-Torregrosa, Role of wettability and nanoroughness on interactions between osteoblast and modified silicon surfaces, Acta Biomater. 7 (2011) 771–778. doi:10.1016/j.actbio.2010.08.024.

[48] S. KENNEDY, N. WASHBURN, C. SIMONJR, E. AMIS, Combinatorial screen of the effect of surface energy on fibronectin-mediated osteoblast adhesion, spreading and proliferation☆, Biomaterials. 27 (2006) 3817–3824. doi:10.1016/j.biomaterials.2006.02.044.

[49] F. Amiri, A. Jahanian-Najafabadi, M.H. Roudkenar, In vitro augmentation of mesenchymal stem cells viability in stressful microenvironments, Cell Stress Chaperones. 20 (2015) 237–251. doi:10.1007/s12192-014-0560-1.

[50] P.O.F. Elsevier, New age Medical clinic, (2007) 1–30.

[51] E.J. Tsang, C.K. Arakawa, P.A. Zuk, B.M. Wu, Osteoblast Interactions Within a Biomimetic Apatite Microenvironment, Ann. Biomed. Eng. 39 (2011) 1186–1200. doi:10.1007/s10439-010-0245-6.

[52] Z. Meleti, I.. Shapiro, C.. Adams, Inorganic phosphate induces apoptosis of osteoblast-like cells in culture, Bone. 27 (2000) 359–366. doi:10.1016/S8756-3282(00)00346-X.

[53] C.S. Adams, K. Mansfield, R.L. Perlot, I.M. Shapiro, Matrix Regulation of Skeletal Cell Apoptosis, J. Biol. Chem. 276 (2001) 20316–20322. doi:10.1074/jbc.M006492200.